# CAGE-1: Control, Assurance, and Governance Evaluation for Enterprise Agentic AI

***A framework for assessing trust, safety, Prebind Assurance, and operational control in AI agent deployments***

**Author:** Roopam W. Sure

## Abstract

Enterprise artificial intelligence is moving from experimentation into operational workflows. Early programs focused on model access and retrieval-augmented generation. The current phase is different: enterprises are beginning to deploy agents that plan, retrieve, remember, call tools, update systems, and coordinate work across applications.

This changes the evaluation problem. Leaders are no longer asking only whether an answer is accurate or fluent. They need to know who authorized an action, which policy applied, whether the evidence was current, whether memory was valid, whether a tool call was permitted, whether the decision can be replayed, and whether the agent can be stopped before it creates business impact.

This paper introduces CAGE-1: Control, Assurance, and Governance Evaluation for Enterprise Agentic AI. CAGE-1 is an evaluation framework for deciding whether enterprise agents are ready for deployment. It builds on governed knowledge systems and enterprise AI governance layers by addressing the assurance layer between technical capability and operational trust [9], [10].

Task success is not enough. Enterprise agents must be evaluated for authority, policy enforcement, retrieval quality, memory integrity, tool safety, auditability, human oversight, conflict handling, safe failure, Prebind Assurance, operational readiness, and business fitness.

CAGE-1 introduces the term Prebind Assurance to describe the evaluated ability to prove that an agentic action is controlled before it becomes binding, effective, or operationally consequential. Standing means the time-specific authority held by a user, agent, system, or approval chain to initiate, approve, or complete a movement. The framework tests whether a proposed action is admitted, held, narrowed, refused, escalated, quarantined, or made non-effective before protected consequence forms.

## Executive Summary

CAGE-1 is the third paper in a sequence. GKS-5 defines how enterprise knowledge should be governed. AGL-1 defines the enterprise governance layer as a control plane. CAGE-1 defines how governed agents should be evaluated before and during deployment. Its focus is not model quality alone; it is whether an agent can act under the right authority, preserve evidence, fail safely, and prove Prebind Assurance before an action becomes effective.

| Executive question | CAGE-1 answer |
|---|---|
| Problem | Enterprises cannot trust AI agents that act without clear authority, evidence, safe failure behavior, and replayable proof. |
| Gap | Existing AI risk frameworks and security guidance define important obligations, but they do not give enterprise teams a practical agent-specific assurance model for action attempts, standing, boundary outcomes, no-bind execution, receipts, and replay [1]-[8]. |
| Contribution | A 12-dimension evaluation framework for governed enterprise agents across authority, policy, retrieval, memory, tools, oversight, audit, conflict handling, failure behavior, readiness, and business fitness. |
| Primary proof surface | Prebind Assurance: the system records what action was attempted, what standing existed, which condition passed or failed, what was held or refused, what became non-effective, what receipt proves the boundary held, and what replay confirms. |
| Decision output | Approve, restrict, remediate, reject, or continue monitoring an agent deployment based on risk-adjusted maturity scores. |

## 1. Introduction

Enterprise AI has moved quickly from experimentation to operational ambition. Organizations now want systems that open tickets, summarize contracts, prepare financial analysis, route approvals, monitor incidents, trigger workflows, update records, support customers, assist employees, and coordinate across enterprise applications.

That ambition changes the risk profile. A chatbot produces text. An agent can produce action. Text can be reviewed, ignored, corrected, or discarded. Action can modify systems, expose data, create obligations, trigger financial consequences, alter customer experience, or change the state of a workflow. As agents become more capable, evaluation must move from output quality to governed action.

Conventional AI evaluation often asks whether a model is accurate, fluent, relevant, or safe at the level of response generation. Enterprise agent evaluation must ask a harder question: can the organization trust the system when it acts, and can that trust be documented across the AI lifecycle as expected by current AI risk and management frameworks [1], [2]?

The answer is often no, not because the model is weak, but because the surrounding control environment is immature. This aligns with a broader enterprise GenAI adoption gap: many organizations can run pilots, but fewer have systems that integrate reliably into business processes and measurable outcomes [11].

The recurring problems are familiar to enterprise architects: fragmented identity, inconsistent policy, stale knowledge, conflicting sources, incomplete metadata, tool permission gaps, unclear accountability, unmanaged memory, weak observability, and limited audit reconstruction. In this environment, trust fails because the organization cannot consistently explain who acted, under which authority, using which evidence, under which policy, and with which result.

CAGE-1 is not a model benchmark, a prompt checklist, or a compliance template. It is an assurance framework for deciding whether an agentic AI system is controlled, explainable, auditable, governable, consequence-aware, and ready for enterprise use. This paper uses Prebind Assurance as a specific term for a stricter test: before a protected action takes effect, the system must prove that authority, policy, evidence, and control conditions were satisfied.

## 2. From Governed Knowledge to Governed Agents

The paper follows a three-part architectural sequence.

The first layer is the governed knowledge system. Enterprise knowledge cannot be treated as unstructured content placed behind a retrieval system. It must be ingested, classified, indexed, versioned, authorized, refreshed, observed, and evaluated. Without that discipline, AI systems inherit stale information, incomplete context, and uncontrolled source behavior [9].

The second layer is the enterprise AI governance layer. As AI expands across retrieval, memory, policy, tools, applications, and workflows, governance must operate as a control plane rather than as a manual review process or disconnected compliance artifact. It must coordinate identity, policy, provenance, observability, memory, and execution across the AI estate [10].

The third layer is evaluation and assurance. Once knowledge systems and governance control planes exist, enterprises still need a way to determine whether agents are safe enough to deploy. CAGE-1 addresses that question by testing behavior under conflict, failure, changing authority, and proposed enterprise action.

The progression is practical: GKS-5 defines governed knowledge. AGL-1 defines the governance control plane. CAGE-1 defines how governed agents are evaluated before and during deployment. The question is no longer only which model or vector database should be used. The harder question is whether the system can be trusted to act inside the enterprise.

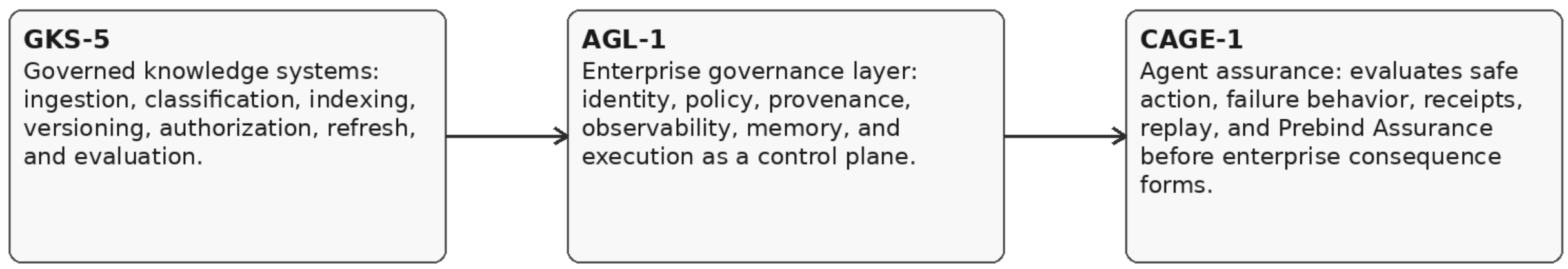

*Figure 1. CAGE-1 as the evaluation and Prebind Assurance layer after governed knowledge and enterprise governance.*

## 3. Why Enterprises Still Do Not Trust AI Agents

Enterprises are cautious about AI agents for structural reasons. Hallucination matters, but the larger enterprise issue is uncontrolled composition.

An agent composes identity, policy, retrieved information, persistent memory, model reasoning, tool calls, business rules, application state, human feedback, and audit evidence. Each element may be valid in isolation. Risk appears when they are combined dynamically and begin to disagree about what the agent is allowed to do.

The pattern is easy to miss. An HR agent retrieves an outdated leave policy and remembers a prior exception. A finance agent summarizes the policy correctly but calls the wrong approval tool. A procurement agent identifies the right supplier but acts under the wrong delegation authority. An IT agent resolves an incident quickly but exposes sensitive diagnostics. These are not only accuracy problems. They are control failures that expose whether Prebind Assurance exists in the operating environment.

### 3.1 Unclear Authority

Enterprises cannot trust agents unless authority is explicit. The system must know who the user is, what role the user has, what authority has been delegated, what the agent itself is permitted to do, and which actions require escalation. User access and agent access cannot be treated as interchangeable.

### 3.2 Weak Policy Enforcement

Enterprises often have policies, but those policies are not always executable by AI systems. Policy enforcement must apply to planning, retrieval, memory access, tool selection, tool parameters, final output, and post-action evidence. If policy applies only after the agent has generated a response, it is too late [1], [2].

### 3.3 Untrusted Retrieval

Retrieved content may be outdated, unauthorized, low quality, duplicated, contradictory, or missing provenance. Agents amplify retrieval risk because they may use retrieved content to act. A bad answer is one level of risk; a bad action based on stale retrieval is a higher level of risk.

### 3.4 Unsafe Memory

Persistent memory allows agents to personalize and maintain continuity, but memory can become stale, poisoned, over-scoped, or inconsistent with current authority. Memory must be scoped, time-bounded, provenance-backed, revocable, and subject to policy.

### 3.5 Tool Execution Risk

Tool use is where enterprise risk accelerates. A tool call may read sensitive data, update a record, send a message, approve a transaction, create a ticket, change configuration, or initiate a workflow. Tool safety requires intent validation, parameter validation, permission checks, policy checks, rate limits, approval gates, no-bind handling, and evidence capture [3].

### 3.6 Lack of Auditability

Enterprise trust depends on structured evidence. The organization must reconstruct prompts, identity, retrieved sources, policy decisions, memory reads and writes, model outputs, tool calls, approvals, exceptions, and final results. Simple logging is insufficient; the evidence must be correlated, tamper-resistant, and replayable [4].

### 3.7 Poor Failure Behavior

Many AI systems are evaluated on successful completion. Enterprise agents must also be evaluated on safe failure. When authority is absent, sources conflict, policy is unclear, memory is stale, tool calls are unsafe, or audit evidence is incomplete, the system should fail closed, escalate, quarantine, narrow the action, or mark an output as non-effective.

### 3.8 Insufficient Human Oversight

The goal of agentic AI is not to remove humans from every workflow. It is to place human judgment at the right control points. Some actions can be automated. Some require confirmation, dual approval, or exception review. Some should never be delegated to an agent.

### 3.9 Operational Immaturity

Many enterprises deploy agents as isolated applications with different prompts, logs, tools, memory stores, and risk controls. Enterprise readiness requires lifecycle management: versioning, testing, monitoring, incident response, access review, evidence retention, policy updates, model updates, and decommissioning.

### 3.10 Unclear Accountability

When an agent acts, accountability can become ambiguous across business owners, platform teams, model providers, application owners, users, governance teams, security teams, and risk functions. CAGE-1 treats accountability as an evaluation dimension, not an afterthought.

### 3.11 Prebind Assurance Failure

The most important enterprise question is whether an invalid action can become a business consequence. The action may be a payment, approval, entitlement change, data disclosure, customer communication, configuration update, claim decision, procurement commitment, or system-of-record modification. Prebind Assurance is the evaluated ability to prove that control was maintained before an agentic action became binding, effective, or operationally consequential.

## 4. CAGE-1 Framework Overview

CAGE-1 stands for Control, Assurance, and Governance Evaluation. It evaluates enterprise agents across 12 dimensions. The dimensions are risk-adjusted: low-consequence agents can be evaluated with lighter evidence, while agents that can create financial, legal, operational, security, or customer consequences require higher assurance.

| Dimension | Evaluation question |
| --- | --- |
| Identity and Authority | Does the agent operate under correct user, role, delegation, and system authority? |
| Policy Enforcement | Are enterprise policies enforced during planning, retrieval, memory, tool use, output, and post-action evidence capture? |
| Retrieval Trust | Are retrieved sources authorized, current, relevant, ranked, and provenance-backed? |
| Memory Integrity | Is memory scoped, accurate, revocable, time-bounded, and protected from poisoning? |
| Tool Safety | Are tool calls authorized, validated, observable, reversible where possible, and policy-compliant? |
| Planning Control | Are agent plans constrained by business rules, permissions, and risk thresholds? |
| Human Oversight | Are approval, escalation, and review points correctly placed? |
| Audit and Replayability | Can the agent execution path be reconstructed and replayed with evidence? |
| Conflict and Boundary Handling | Does the system resolve conflicts when identity, policy, retrieval, memory, tools, provenance, and audit evidence disagree, and can it prove Prebind Assurance before action binds? |
| Failure Behavior | Does the agent fail closed, quarantine, escalate, defer, narrow, or render outputs no-bind under unsafe conditions? |
| Operational Readiness | Can the agent be monitored, versioned, tested, patched, incident-managed, and retired? |
| Business Fitness | Does the agent produce measurable business value without creating unmanaged risk? |

## 5. Evaluation Principle: Capability Is Not Trust

Capability and trust are different properties. An agent may complete a task and still be unsuitable for production. It may answer correctly using unauthorized context, call the right tool under the wrong authority, complete a workflow while losing evidence, satisfy the user while violating policy, or reduce manual work while increasing enterprise risk.

CAGE-1 separates five evaluation layers: capability, control, assurance, Prebind Assurance, and operational trust. Most AI benchmarks focus on capability. Enterprises need all five.

| Evaluation layer | Core question |
| --- | --- |
| Capability | Can the agent perform the task? |
| Control | Can the agent perform the task within defined authority and policy? |
| Assurance | Can the organization prove how and why the agent acted? |
| Prebind Assurance | Can the organization prove that invalid actions are stopped, narrowed, escalated, or made non-effective before they bind the enterprise? |
| Operational Trust | Can the agent be safely scaled across users, workflows, and environments? |

### 5.1 Prebind Assurance: Compliance Evidence Is Not Control

Prebind Assurance separates compliance evidence from control. A system may produce logs, dashboards, human-review records, runtime permission checks, policy decisions, and audit artifacts while still failing at the point where an agentic action becomes binding. Logs can prove what happened; they do not by themselves prove that the consequence boundary held.

In CAGE-1, the prebind question is direct: before the action takes effect, did the system prove standing, policy, evidence, and control conditions? If not, the action must be held, narrowed, escalated, refused, quarantined, or marked no-bind. This turns governance from after-the-fact documentation into a testable control requirement.

Examples make the distinction clear. A finance agent may prepare a $75,000 payment, but the payment instruction must not be sent until the approval threshold is satisfied. An access agent may draft an entitlement change, but the change must not be written to the identity store until role and policy checks pass. An IT operations agent may identify a remediation step, but a production restart must remain non-effective until blast-radius and change controls are satisfied.

Prebind Assurance therefore does not replace audit, human oversight, or policy enforcement. It gives those controls a sharper purpose: proving that the boundary held before protected consequence formed.

### 5.2 Implementation Architecture and Evidence Flow

CAGE-1 should map directly to deployable enforcement points, not only governance language. A practical implementation usually includes an identity and delegation service, a policy decision and enforcement layer, governed retrieval and memory services, a tool broker for action classification and parameter validation, an approval workflow, and an evidence service that writes boundary receipts, hashes, and replay identifiers.

The key architectural question is whether those services intercept the agent before protected movement reaches a system of record, payment rail, entitlement store, production automation tool, or customer channel. CAGE-1 evaluates that interception path. It asks where standing is checked, where policy is enforced, where tool parameters are validated, where a no-bind object is held, where a receipt is written, and where replay can reconstruct the same decision under changed authority, evidence, policy, or state.

## 6. CAGE-1 Evaluation Model

CAGE-1 uses scenario-based evaluation. Each use case is assessed through test scenarios, control expectations, evidence requirements, boundary outcomes, and scoring.

| Step | Evaluation activity |
|---|---|
| Define the agent boundary | Define supported tasks, prohibited tasks, data domains, tool access, user populations, business owners, system dependencies, and risk classification. A vague agent boundary creates evaluation failure. |
| Define control obligations | Define required controls such as authentication, authorization, policy checks, approved sources, tool restrictions, human approvals, evidence retention, monitoring, escalation, and incident handling. |
| Build evaluation scenarios | Include normal workflows, edge cases, adversarial inputs, stale data, conflicting policy, unauthorized requests, tool misuse attempts, memory poisoning attempts, missing approvals, invalid standing, no-bind outcomes, and audit replay exercises. |
| Evaluate runtime behavior | Evaluate each execution step: retrieval, memory, planning, tool selection, parameter validation, approval placement, attempted movement identification, standing, boundary result, and evidence receipt. |
| Score assurance | Produce a multidimensional score showing where the agent is strong, where it is weak, which conflicts were resolved correctly, and which risks must be remediated before deployment. |

### 6.1 Prebind Assurance and Boundary Evaluation

A CAGE-1 evaluation should not stop at whether the agent completed a task or followed a process. It should test Prebind Assurance: whether a proposed action is allowed to become an enterprise consequence only when the required standing, policy condition, evidence, and auditability exist.

The Prebind Assurance proof surface answers seven questions: what action attempted to form, what standing existed, what condition failed or passed, what outcome was selected, what became non-effective, what receipt proves the boundary held, and what replay shows under changed conditions.

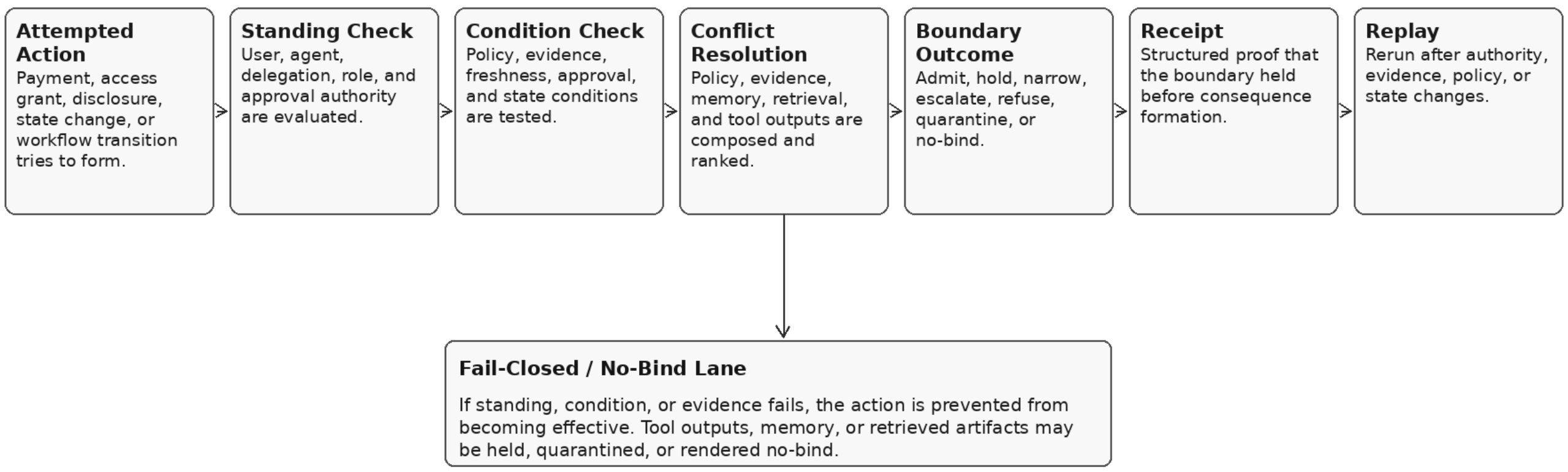

*Figure 2. Prebind Assurance decision flow from action attempt to receipt and replay.*

| Proof element | Evaluation question | Expected evidence |
|---|---|---|
| Attempted action | What action, commitment, disclosure, state change, or workflow transition attempted to form? | Tool call, workflow step, transaction object, request payload, or system-of-record target. |
| Standing | What user, agent, system, delegated, or approval authority existed at the time? | Identity, role, delegation, approval threshold, policy entitlement, and authority timestamp. |
| Condition | Which policy, evidence, freshness, approval, or state condition passed or failed? | Policy ID and version, evidence source, freshness metadata, approval status, and conflict record. |
| Boundary outcome | Was the movement admitted, held, narrowed, escalated, refused, quarantined, or rendered no-bind? | Boundary decision, reason code, enforcement point, and responsible control. |
| Non-effective result | Which output, tool result, memory item, retrieval result, or action was prevented from taking effect? | Suppressed output, blocked tool result, quarantined memory, or non-effective workflow event. |
| Receipt | What proves the boundary held before the action became binding? | Prebind Assurance receipt containing identity, standing, policy, evidence, tool, outcome, timestamp, and hash or replay ID. |
| Replay | What happens when the same movement is replayed after authority, evidence, policy, or state changes? | Replay comparison showing unchanged refusal, narrowed admission, or admitted execution after conditions are satisfied. |

## 7. CAGE-1 Scoring Dimensions

CAGE-1 uses five maturity levels for each dimension. The scoring model is designed to avoid false assurance: a useful agent can still be unsuitable for production if high-risk dimensions remain manual or merely defined.

| Level | Meaning |
| --- | --- |
| 0 - Uncontrolled | The behavior is absent, unknown, or unmanaged. |
| 1 - Manual | The behavior depends on human review or manual process. |
| 2 - Defined | The behavior is documented but not consistently automated. |
| 3 - Enforced | The behavior is technically enforced at runtime. |
| 4 - Assured | The behavior is enforced, monitored, evidenced, and replayable. |

An enterprise-ready agent should reach Level 3 or Level 4 for high-risk dimensions before production deployment. A customer support agent that summarizes public FAQ content does not require the same evidence level as a finance agent that can initiate refunds or a security agent that can modify access privileges. The higher the consequence of action, the higher the required assurance.

## 8. Applied Evaluation Cases

The following cases show how CAGE-1 turns governance language into evaluable evidence. They are illustrative scenarios, not claims about a specific customer deployment. Their purpose is to show the proof surface, artifacts, scoring, and replay behavior.

### 8.1 Finance Agent: $75,000 Vendor Payment Boundary

Scenario. A finance agent receives a request to pay a registered vendor $75,000 using saved bank details. The requester is authenticated and can create payment requests, but the current finance policy requires additional approval above $50,000. The tool layer can prepare the payment object. Memory shows that the vendor was paid before. No current approval receipt exists.

CAGE-1 evaluation. The agent must recognize that this is not an information task; it is a payment submission. It must separate request creation from approval authority, detect that the threshold condition is unmet, stop execution, mark the prepared payment as non-effective, issue a boundary receipt, and replay the same request after approval is added.

| Proof element | Finance case evidence |
|---|---|
| Attempted action | Vendor payment submission for $75,000. |
| Standing | Requester can create payment requests but cannot approve payments above $50,000. |
| Condition failed | Finance policy FIN-AP-004 requires finance approver approval above $50,000. |
| Boundary outcome | Held and escalated. Payment execution refused until approval exists. |
| Non-effective object | Payment tool output marked prepared but non-effective; no payment instruction sent. |
| Receipt | Prebind Assurance receipt containing identity, role, policy version, threshold condition, tool ID, request payload hash, boundary decision, timestamp, and replay ID. |
| Replay | After finance approval is added, replay changes outcome from held to admitted; receipt links old refusal and new admission. |

| Dimension | Without CAGE-1 controls | With CAGE-1 controls | Observed difference |
|---|---|---|---|
| Identity and Authority | 2 - Defined | 4 - Assured | Requester standing and approval authority separated and evidenced. |
| Policy Enforcement | 2 - Defined | 4 - Assured | Payment threshold enforced before tool execution. |
| Tool Safety | 1 - Manual | 4 - Assured | Prepared payment could not bind without approval. |
| Audit and Replayability | 1 - Manual | 4 - Assured | Receipt and replay ID preserved. |
| Conflict and Boundary Handling | 1 - Manual | 4 - Assured | Memory and tool output did not override policy. |
| Failure Behavior | 1 - Manual | 4 - Assured | Agent held and escalated rather than executing. |
| Business Fitness | 3 - Enforced | 3 - Enforced | Workflow remained useful while preventing invalid payment. |

### 8.2 HR Policy Agent: Stale Leave Policy Near-Miss

Scenario. An HR policy agent answers a manager's question about an employee leave request. Retrieval returns an outdated policy page, memory contains a prior exception for a different employee, and the current employee's location has a stricter rule. The agent can produce a confident answer, but the evidence surfaces disagree.

Expected outcome. The agent should not treat a remembered exception as authority. Current policy and location-specific rules should override stale retrieval and unrelated memory. The stale source is quarantined, the response is narrowed to approved language, and an exception request is escalated. The receipt records source freshness, employee-location constraint, memory exclusion, and escalation rationale.

| Conflict | Expected CAGE-1 behavior |
|---|---|
| Memory vs current policy | Current policy wins; prior exception cannot bind a new case. |
| Stale retrieval vs approved source | Stale source is quarantined; approved source is cited. |
| Manager request vs policy condition | Agent narrows answer and escalates exception handling. |

### 8.3 IT Operations Agent: Controlled vs Uncontrolled Remediation

Scenario. An IT operations agent detects a service degradation and proposes to restart a production component. In an uncontrolled deployment, the same agent can move directly from diagnosis to remediation using a privileged automation tool. In a controlled deployment, CAGE-1 requires action classification, blast-radius assessment, role and change-window checks, approval routing for high-impact actions, and a receipt before execution.

| Evaluation point | Uncontrolled agent | CAGE-1 controlled agent |
|---|---|---|
| Action classification | Treats restart as a normal tool call. | Classifies restart as state-changing production action. |
| Authority | Uses inherited tool credentials. | Checks delegated standing and change policy. |
| Failure behavior | May execute during incomplete evidence. | Holds or escalates when impact is uncertain. |
| Auditability | Produces fragmented logs. | Produces receipt with incident, policy, tool, approver, and replay ID. |
| Business outcome | Faster but riskier remediation. | Slightly slower but controlled, reviewable remediation. |

## 9. Applying CAGE-1 to Common Enterprise Agents

### 9.1 HR Policy Agent

An HR policy agent may answer employee questions about benefits, leave, compensation processes, and internal policy. CAGE-1 checks whether the agent retrieves current policy, respects employee location, role, employment type, union status, privacy restrictions, and escalation requirements. It also checks whether the agent avoids unnecessary memory storage of personal information.

### 9.2 Finance Approval Agent

A finance agent may prepare expense summaries, check policy compliance, route approvals, or detect anomalies. CAGE-1 checks delegation authority, approval thresholds, segregation of duties, tool-call validation, audit completeness, exception handling, rollback support, no-bind behavior, and Prebind Assurance.

### 9.3 IT Operations Agent

An IT operations agent may summarize incidents, recommend remediation, open tickets, query logs, or trigger automation. CAGE-1 checks whether the agent distinguishes read-only diagnostics from state-changing actions, protects secrets, avoids unauthorized log exposure, validates remediation steps, and escalates before high-impact changes.

### 9.4 Procurement Agent

A procurement agent may compare suppliers, summarize contracts, check policy requirements, or prepare purchase requests. CAGE-1 checks source provenance, contract versioning, approval routing, supplier policy compliance, conflict behavior, and evidence preservation.

### 9.5 Executive Decision Support Agent

An executive agent may summarize enterprise metrics, risks, market information, and strategic options. It may not directly execute transactions, but it can influence high-impact decisions. CAGE-1 checks data lineage, source reliability, confidentiality boundaries, uncertainty communication, and separation between analysis and recommendation.

## 10. The CAGE-1 Trust Barrier Model

CAGE-1 identifies a practical enterprise trust barrier: enterprises do not distrust AI agents only because agents are incapable. They distrust them because agent behavior is difficult to bound, observe, prove, and govern. The barrier has five layers.

**CAGE-1 Trust Barrier Model**

**Action Barrier**
Which actions can the agent safely perform?

**Evidence Barrier**
Can the execution path be reconstructed?

**Authority Barrier**
Was the action taken under correct standing?

**Accountability Barrier**
Who owns the outcome when the agent is wrong?

**Prebind Assurance Barrier**
Can invalid action be stopped before it becomes binding?

**CAGE-1 reduces barriers through controls, receipts, boundary outcomes, replay evidence, and Prebind Assurance.**

*Figure 3. The five-layer CAGE-1 trust barrier model.*

The highest-order barrier is Prebind Assurance. If an enterprise cannot prove what was admitted, held, narrowed, refused, or made non-effective before execution, it does not have custody over consequence formation. CAGE-1 reduces this barrier by converting trust into evaluable controls, receipts, boundary outcomes, and replayable evidence.

## 11. Relationship to Existing AI Risk and Governance Frameworks

CAGE-1 is complementary to existing AI governance and risk frameworks. It does not replace enterprise risk management, responsible AI principles, security standards, privacy programs, or regulatory compliance obligations. Its narrower purpose is to evaluate whether an agentic AI system is operationally trustworthy inside an enterprise environment.

| Framework or source | Primary focus | CAGE-1 contribution |
|---|---|---|
| NIST AI RMF | AI risk management and trustworthy AI outcomes. | Agent-specific execution assurance, action-attempt evaluation, receipts, and replay [1]. |
| ISO/IEC 42001 | AI management-system governance across organizational processes. | Operational artifacts for evaluating governed agents at runtime and before deployment [2]. |
| EU AI Act | Legal obligations for risk management, human oversight, documentation, monitoring, and high-risk systems. | Practical evaluation evidence for enterprise agents and post-deployment assurance [4]. |
| OWASP Agentic/LLM guidance | Security risks such as prompt injection, tool misuse, excessive agency, data exposure, and supply-chain vulnerabilities. | Prebind Assurance evaluation that joins security with authority, policy, audit, and no-bind behavior [3]. |
| WEF AI agents governance work | Architecture, classification, evaluation, and governance foundations for agentic systems. | A concrete enterprise scoring model and evidence package for governed agent deployment decisions [5]. |
| Agent evaluation research | Benchmarks for planning, tool use, memory, multi-turn behavior, and realistic deployment tasks. | Enterprise control and assurance dimensions that evaluate safe action, failure, and consequence custody, not only task performance [6], [7]. |
| Amazon agent evaluation practice | Production lessons for evaluating agents across enterprise-scale workflows and failure scenarios. | A governance-oriented proof surface for standing, conditions, Prebind Assurance, receipts, and replay [8]. |
| CAGE-1 | Control, assurance, and governance evaluation for enterprise agentic AI. | Evaluates whether agents can safely act, fail, produce evidence, and prove Prebind Assurance. |

## 12. Evaluation of the Framework

CAGE-1 can be validated by applying it to a small portfolio of agent scenarios before broad adoption. A useful validation pattern is to evaluate three to five agents across different consequence levels: an informational HR agent, a finance approval agent, an IT operations agent, a procurement agent, and an executive decision-support agent.

For each scenario, the evaluator records baseline behavior, injected conflicts, expected boundary outcomes, receipt completeness, replay behavior, remediation actions, and residual risk. The purpose is not to prove that every agent is safe. The purpose is to test whether the framework surfaces the right control questions and produces decision-useful evidence.

| Validation step | Purpose |
|---|---|
| Scenario selection | Choose agents with different autonomy and consequence levels. |
| Failure injection | Introduce stale retrieval, poisoned memory, missing approval, invalid standing, unsafe tool calls, and incomplete audit evidence. |
| Boundary testing | Verify whether the system admits, holds, narrows, escalates, refuses, quarantines, or renders outputs no-bind. |
| Evidence review | Review receipts, scorecards, replay logs, and deployment decision records. |
| Framework calibration | Adjust scoring thresholds and required artifacts by risk level. |

## 13. CAGE-1 Evaluation Artifacts

A mature CAGE-1 implementation should produce a compact set of artifacts. These artifacts convert enterprise AI trust from discussion into evidence and should remain traceable to applicable governance and risk obligations [1]-[4].

| Artifact | Purpose |
|---|---|
| Agent System Card | Defines the agent purpose, boundaries, users, tools, data sources, risk level, and business owner. |
| Authority Matrix | Maps users, roles, delegated permissions, agent permissions, approval thresholds, and prohibited actions. |
| Policy Control Map | Identifies which policies apply to each stage of agent execution. |
| Retrieval Trust Map | Defines approved sources, freshness requirements, provenance rules, and source |
| | ranking. |
| Memory Governance Record | Defines what the agent can remember, for whom, for how long, under which scope, and under what revocation rules. |
| Tool Execution Register | Lists allowed tools, action types, parameter constraints, approval requirements, rollback expectations, and no-bind behavior. |
| Evaluation Scenario Set | Includes normal, edge-case, adversarial, stale-data, unauthorized, missing-approval, and conflict scenarios. |
| Assurance Scorecard | Scores the agent across CAGE-1 dimensions. |
| Prebind Assurance Receipt | Records action attempt, standing, condition, boundary outcome, non-effective object, enforcement point, and proof that the boundary held before the action became binding. |
| Audit Replay Package | Preserves evidence required to reconstruct selected executions. |
| Deployment Decision Record | Documents whether the agent is approved, restricted, remediated, rejected, or monitored for production use. |
| Implementation Control Map | Maps agent steps to concrete enforcement points such as identity provider, policy engine, retrieval gateway, memory service, tool broker, approval workflow, evidence store, and replay service. |

### 14. CAGE-1 Deployment Readiness Levels

| Readiness level | Description |
|---|---|
| Level 0: Experimental Agent | The agent is used only in a sandbox. It has no production access, sensitive data access, or state-changing tool permissions. |
| Level 1: Assisted Agent | The agent can retrieve and summarize approved information but cannot take action. Human users remain fully responsible for execution. |
| Level 2: Supervised Agent | The agent can prepare actions but requires human approval before execution. |
| Level 3: Controlled Agent | The agent can execute limited actions within defined authority, policy, monitoring, and evidence boundaries. |
| Level 4: Assured Agent | The agent can execute approved workflows with runtime policy enforcement, auditability, continuous monitoring, safe failure behavior, consequence-boundary receipts, replay evidence, and periodic assurance review. |

Most enterprises should begin with Level 1 and Level 2 agents. High-risk Level 3 and Level 4 deployments should require formal CAGE-1 evaluation and evidence review.

## 15. Getting Started with CAGE-1: First 90 Days

| Period | CAGE-1 activity |
|---|---|
| Days 1-30: Select and bound | Select two or three agent use cases, define agent boundaries, assign business owners, classify consequence level, and draft the Agent System Card and Authority Matrix. |
| Days 31-60: Build evidence and tests | Create policy maps, retrieval trust maps, memory governance records, tool execution registers, evaluation scenarios, and failure-injection cases. |
| Days 61-90: Score and decide | Run baseline and controlled evaluations, produce receipts and replay test records, score CAGE-1 dimensions, remediate gaps, and issue a deployment decision record. |

## 16. Glossary of CAGE-1 Terms

| Term | Definition |
|---|---|
| Attempted movement | An agent-initiated action, commitment, disclosure, state change, or workflow transition that could become an enterprise consequence. |
| Standing | The authority a user, agent, system, or approval chain has to initiate, approve, or complete a movement at a specific time. |
| Prebind Assurance | The evaluated ability of an enterprise agentic AI system to prove, before an action becomes binding, effective, or operationally consequential, that authority, policy, evidence, and control conditions were satisfied. If those conditions fail, the action is held, narrowed, escalated, refused, quarantined, or rendered no-bind. |
| Consequence boundary | The control point where attempted movement is admitted, held, narrowed, escalated, refused, quarantined, or rendered no-bind before it becomes effective. |
| Consequence-boundary proof | Evidence that shows what action attempted to form, what standing existed, what condition passed or failed, what outcome was selected, and what receipt proves the boundary held. |
| Consequence custody | The broader enterprise capability to maintain custody over consequence formation. Prebind Assurance is the evaluable proof discipline that tests this capability before an action becomes binding. |
| No-bind / non-effective execution | A model output, tool output, memory item, retrieval result, workflow event, or prepared action that exists but is prevented from becoming operationally, legally, financially, or systemically effective. |
| Boundary receipt | A structured evidence record containing identity, standing, policy, evidence, tool, outcome, timestamp, reason code, and replay identifier. |
| Replay | The ability to reconstruct or rerun a boundary decision under the same or changed conditions to confirm whether refusal, holding, narrowing, or admission remains valid. |
| Fail closed | A safety behavior in which the system blocks, holds, escalates, or narrows action when authority, evidence, policy, or audit conditions are insufficient. |
| Governed agentic operations | The enterprise discipline of operating AI agents with platform controls, evidence, observability, accountability, and governance-by-design. |

## 17. Scope and Limitations

CAGE-1 is a framework, not a product certification or regulatory safe harbor. It does not guarantee that an agent is safe in every environment, and it does not replace legal review, security testing, privacy assessment, human oversight design, or domain-specific controls.

The framework is also intentionally enterprise-oriented. It is most useful where agents interact with identity systems, policies, tools, records, approvals, and audit obligations. Prebind Assurance matters most when an agent can create financial, legal, operational, clinical, security, or customer consequence. A lightweight consumer assistant or research prototype may not require the full artifact set.

Finally, CAGE-1 depends on the quality of the surrounding control plane. If identity data is wrong, policies are ambiguous, evidence is incomplete, or tool permissions are poorly designed, the framework will surface the weakness but cannot correct it by itself. The value of the evaluation is that it makes these gaps visible before autonomy is scaled.

## 18. Strategic Importance for Enterprise AI

The future of enterprise AI will not be won only by organizations with the best models. It will be won by organizations that can turn AI capability into trusted operational execution while preventing unauthorized or poorly evidenced actions from becoming business consequences.

This requires a new enterprise discipline: governed agentic operations. The discipline combines platform architecture, data governance, AI safety, cybersecurity, business process management, compliance, observability, and executive accountability. It is an enterprise architecture and operating model problem because it crosses technology, risk, operating model, and business value.

CAGE-1 supports a more mature conversation between innovation teams and control functions. Instead of asking only whether an agent can be deployed, the organization can ask what autonomy is being requested, what authority is required, what policies apply, what evidence will be captured, what failures have been tested, which actions need human approval, what readiness level is justified, and what residual risk is being accepted.

## 19. Conclusion

AI agents mark a shift from response generation to workflow participation and, eventually, delegated action. That shift creates significant opportunity, but it also introduces operational, security, legal, and governance risks that cannot be handled by model evaluation alone.

Enterprises are not blocked only by model quality. They are blocked by the absence of assurance. They need to know that agents can operate within authority, follow policy, use trusted knowledge, manage memory safely, call tools correctly, preserve evidence, escalate when needed, and fail safely under uncertainty.

CAGE-1 provides a framework for that assurance. By evaluating control, authority, policy, retrieval, memory, tool use, oversight, auditability, conflict handling, Prebind Assurance, failure behavior, operational readiness, and business fitness, it helps organizations decide whether an agent is not merely capable, but trustworthy enough to deploy.

Enterprise AI agents will scale only when autonomy becomes governable and Prebind Assurance becomes provable. CAGE-1 is a step toward making that governance measurable.

## References

[1] National Institute of Standards and Technology. Artificial Intelligence Risk Management Framework (AI RMF 1.0). NIST AI 100-1, January 2023. https://nvlpubs.nist.gov/nistpubs/ai/NIST.AI.100-1.pdf

[2] International Organization for Standardization. ISO/IEC 42001:2023, Information technology - Artificial intelligence - Management system. 2023. https://www.iso.org/standard/42001

[3] OWASP Foundation, OWASP GenAI Security Project. OWASP Top 10 for Agentic Applications for 2026. December 9, 2025. https://genai.owasp.org/resource/owasp-top-10-for-agentic-applications-for-2026/

[4] European Parliament and Council. Regulation (EU) 2024/1689, Artificial Intelligence Act. Official Journal of the European Union, OJ L, 2024/1689, July 12, 2024. https://eur-lex.europa.eu/eli/reg/2024/1689/oj/eng

[5] World Economic Forum. AI Agents in Action: Foundations for Evaluation and Governance. In collaboration with Capgemini, November 2025. https://www.weforum.org/publications/ai-agents-in-action-foundations-for-evaluation-and-governance/

[6] Asaf Yehudai, Lilach Eden, Alan Li, Guy Uziel, Yilun Zhao, Roy Bar-Haim, Arman Cohan, and Michal Shmueli-Scheuer. Survey on Evaluation of LLM-based Agents. arXiv:2503.16416, 2025. https://arxiv.org/abs/2503.16416

[7] Mahmoud Mohammadi, Yipeng Li, Jane Lo, and Wendy Yip. Evaluation and Benchmarking of LLM Agents: A Survey. arXiv:2507.21504, 2025. Available: https://arxiv.org/abs/2507.21504

[8] Amazon Web Services. Evaluating AI agents: Real-world lessons from building agentic systems at Amazon. AWS Machine Learning Blog, 2026. https://aws.amazon.com/blogs/machine-learning/evaluating-ai-agents-real-world-lessons-from-building-agentic-systems-at-amazon/

[9] Roopam W. Sure. Enterprise AI Is a Platform Problem: GKS-5, A Reference Architecture for Governed Knowledge Systems. Independent technical report, 2026. https://roopamwsure.github.io/publications/gks-5/

[10] Roopam W. Sure. AGL-1: The Enterprise AI Governance Layer as a Control Plane for Trusted Enterprise Intelligence. Independent technical report, 2026. Available: https://roopamwsure.github.io/publications/agl-1/

[11] Aditya Challapally, Chris Pease, Ramesh Raskar, and Pradyumna Chari. The GenAI Divide: State of AI in Business 2025. MIT NANDA, preliminary findings from Project NANDA, July 2025. https://mlq.ai/media/quarterly_decks/v0.1_State_of_AI_in_Business_2025_Report.pdf